\documentclass[journal]{IEEEtran}
\usepackage{amssymb}
\usepackage{amsmath}
\usepackage[caption=true]{caption}
\usepackage{epsfig}
\usepackage{amsthm}
\usepackage{mathtools}
\usepackage{multirow}
\usepackage{float}
\usepackage{color}
\usepackage{cite}
\usepackage{graphicx}\usepackage{graphics}
\usepackage{epstopdf}
\usepackage{etoolbox}

\floatstyle{plain}
\newfloat{bottomfigure}{bp}{lop}
\newtheorem{theorem}{Theorem}

\newtheorem{lemma}{Lemma}
\newtheorem{definition}{Definition}
\newtheorem{assumption}{Assumption}
\newtheorem{remark}{Remark}
\newtheorem{problem}{Problem}

\def\bbeta{\text{$\boldsymbol{\eta}$}}

\def\bgamma{\text{$\boldsymbol{\gamma}$}}

\def\bGamma{\text{$\boldsymbol{\Gamma}$}}

\def\b0{\text{$\mathbf{0}$}}
\def\bone{\text{$\mathbf{1}$}}

 	\def\bA{\text{$\mathbf{A}$}}

 	\def\bB{\text{$\mathbf{B}$}}

 	\def\bC{\text{$\mathbf{C}$}}

\def\be{\text{$\mathbf{e}$}} 	
\def\dbe{\text{$\dot{\mathbf{e}}$}}

\def\bff{\text{$\mathbf{f}$}} 	\def\bF{\text{$\mathbf{F}$}}

 	\def\bG{\text{$\mathbf{G}$}}

 	\def\bH{\text{$\mathbf{H}$}}
\def\bHs{\text{$\mathbf{H}^{\scriptsize (s)}$}}

	 	\def\bI{\text{$\mathbf{I}$}}

 		\def\bJ{\text{$\mathbf{J}$}}

 	\def\bK{\text{$\mathbf{K}$}}

 		\def\bL{\text{$\mathbf{L}$}}

 	\def\bM{\text{$\mathbf{M}$}}

 	\def\bP{\text{$\mathbf{P}$}}
\def\cP{\text{${\mathcal{P}}$}}

\def\bq{\text{$\mathbf{q}$}} 	\def\bQ{\text{$\mathbf{Q}$}}

 	\def\bR{\text{$\mathbf{R}$}}
\def\bbR{\mathbb{{R}}}

\def\bs{\text{$\mathbf{s}$}} 	
\def\dbs{\text{$\dot{\mathbf{s}}$}}

\def\bu{\text{$\mathbf{u}$}}

\def\bx{\text{$\mathbf{x}$}} 	
\def\dbx{\text{$\dot{\mathbf{x}}$}}

\def\by{\text{$\mathbf{y}$}} 	\def\bY{\text{$\mathbf{Y}$}}

 	\def\bZ{\text{$\mathbf{Z}$}}

\def\tgamma{\mbox{$\tilde{\bgamma}$}}
\def\hgamma{\mbox{$\hat{\bgamma}$}}

\def\diag{\text{diag}}

\def\bHs{\mbox{{\bf H}}^{\scriptsize\mbox{(s)}}}

\begin{document}
\title{Synchronization in Networked Systems with Parameter Mismatch: Adaptive Decentralized and  Distributed Controls}
\author{Saeed Manaffam$^\dag$, Alireza Seyedi$^{\dagger\dagger}$, Azadeh Vosoughi$^\dag$, {\em Senior Member}, and Tara Javidi$^{\dagger\dagger\dagger}$, {\em Senior Member}% <-this % stops a space
\thanks{$^{\dag}$ The authors are with the Department of Electrical Engineering and Computer Sciences, University of Central Florida, Orlando, FL, emails: {\tt\small saeedmanaffam@knights.ucf.edu, azadeh@ucf.edu.}}
\thanks{$^{\dagger\dag}$ The author was with the Department of Electrical Engineering and Computer Sciences, University of Central Florida, Orlando, FL.}
\thanks{$^{\dag\dag\dag}$ The author is with the Department of Electrical and Computer Engineering, University of California at San Diego, San Diego, CA, email: {\tt\small tjavidi@ece.ucsd.edu}.}
}

\maketitle
\date{\today}

%% -------------------------SECTION (I)        ABSTRACT-------------------
\begin{abstract}
	Here, we study the ultimately bounded stability of network of mismatched systems using Lyapunov direct method. We derive an 
	upper bound on the norm of the error of network states from its average states, which it achieves in finite time. Then, we devise a decentralized compensator to asymptotically pin the network of mismatched systems to a desired trajectory. Next, we  design distributed estimators to compensate for the mismatched parameters performances of adaptive decentralized and distributed compensations are analyzed. Our analytical results are verified by several simulations in a network of globally connected Lorenz oscillators.
\end{abstract}

\textbf{\textit{Index terms}:} Synchronization, complex networks, adaptive control, pinning control, distributed control, parameter mismatch.

%%%%%%%%%%%%%%%%%%%%%%%%%%%%%%%%
%%%%%%-----------------INTRODUCTION
%%%%%%%%%%%%%%%%%%%%%%%%%%%%%%%%
\section{Introduction}

The study of collective behavior of connected systems has drawn significant amount of attention in variety of disciplines spanning from theoretical sciences to engineering. As the study found more applications in smart grids, biological systems, etc. more scientists put effort to understand and solve the related problems in such systems \cite{Wiener65, Winfree67, Pecora98, Chen07, Manaffam.ACC13, Yu09, DeLellis11, Manaffam.TCAS13}. % -\textcolor{red}{\textbf{add recent publication}}.
Synchronization in the networked systems as one manifestation of such collective behavior was first introduced by Wiener \cite{Wiener65}. Pursued by Winfree in his pioneering work \cite{Winfree67}, the problem of synchronization of network of identical system was recognized as relevant in many fields of research \cite{Pecora98, Pirani.TAC16, Chen07, Yu09, DeLellis11, Manaffam.TCAS13, Manaffam.ACC13}. Introduction of master stability function framework by Pecora and Carroll \cite{Pecora98}, made it possible to separate the topological impact of the network from the dynamical properties of individual nodes on synchronizability of the networked systems \cite{Pecora98}. This type of master stability function uses Lyapunov exponents as a measure of stability which provides necessary conditions on the stability of the network \cite{Pecora98}. However, most recently, the sufficient conditions the stability of the network has been provided by Lyapunov direct method and assuming that the nonlinear systems satisfy a very general assumption, namely, Quad$-$condition or its equivalents \cite{Manaffam.ACC13, Yu09, DeLellis11}. 

As experimental studies have shown, similar to the network of identical systems, the network of semi-similar systems also exhibits certain collective behaviors \cite{Restrepo04, Sun09, Sorrentino11,Acharyya12,Manaffam.IET16}. The semi-similarity in these studies implies identical structure for systems while the parameters of the systems in the network can slightly differ from one another \cite{Restrepo04, Sun09, Sorrentino11, Acharyya12, Manaffam.IET16}. In \cite{Restrepo04}, it is reported that if in the network of semi-similar systems, if the parameters of couplings and isolated systems are slightly different, the states of all the system although cannot be absolutely synchronized, however, they can approach to a close vicinity of each other as the states evolve. The results of this work have been provided mostly on experimental merit. Following \cite{Restrepo04} and similar experimental works, a sensitivity analysis for mismatch systems and concept of $\varepsilon$-synchronization have been given in \cite{Sun09, Sorrentino11, Acharyya12}. In \cite{Sun09}, by assuming the parameter mismatch only in isolated systems, an approximate master stability function for the radius of the neighborhood which the trajectories in the network converges has been calculated. The results in \cite{Sun09} are generalized by \cite{Sorrentino11} by introducing mismatches in the inner coupling as well as weights of the connections. In \cite{Acharyya12}, a new master stability function is given by including higher terms in Taylor series of states around the average trajectories of the network. Additionally, coupling optimization to achieve ``best synchronization properties have been given \cite{Acharyya12}. The results of previous work are generalized in \cite{Manaffam.IET16} for weighted directed systems where it has shown that the center of the neighborhood for the trajectories is the weighted average of trajectories where the weights belong to left null space of the Laplacian matrix of the network. For symmetric networks, this weighted average reduces to simple average as assumed by \cite{Restrepo04, Sun09, Sorrentino11, Acharyya12}. Also probability of $\varepsilon$-stability is used as a measure to study the phase transition of the network from desynchronization to $\varepsilon$-synchronization \cite{Manaffam.IET16}. 

As the analysis of the synchronization in networks of the identical systems with no regulations seems to be advanced, the problem of \textit{pinning} has emerged in various fields of researches \cite{ Chen07}\cite{Manaffam.ACC13}\cite{ Yu09}\cite{Pirani.TAC16}. The objective of pinning is to have the network synchronize to a reference trajectory/state, where the reference trajectory is only available in fraction of the locations in the network, where objective of the problem is to locate the systems which would stabilizes the netwok by providing the reference in minimal number of pinned nodes \cite{Porfiri06}. This problem has been studied in many literature such as \cite{ Chen07}\cite{Manaffam.ACC13}\cite{ Yu09}\cite{Pirani.TAC16} \cite{Bapat.Cyb16}\cite{ DeLellis13} and references therein. This method also has been used in cooperative control schemes where the network is spatially distributed and providing the reference trajectory to all the systems is not desirable \cite{DeLellis13, Manaffam.TCST16}.

In this paper, first, we investigate the problem of $\varepsilon-$synchronization in the symmetric network of mismatched oscillators. Using Lyapunov direct method, we find an upper bound on the error of trajectories from the average of trajectories, where the network converges in finite time. The stated conditions on achieving $\varepsilon-$synchronization in finite time are sufficient and it also applies to time varying mismatches. Note that the bounds given in \cite{Sun09, Sorrentino11,Acharyya12,Manaffam.IET16} are asymptotic bounds and only true for constant parameter mismatches not time varying ones. Then, we devise decentralized and distributed mismatch estimators to compensate for the parameter mismatches of the oscillators. It is shown that if in decentralized control method, the reference trajectory is provided for all the systems, the network of mismatched systems can asymptotically converge to the reference. Since, in most applications, the availability of reference model and/or trajectory in all locations is not desirable, we cooperative/distributed control via pinning to synchronize the mismatched network to the reference trajectory. In the distributed scheme, we assume that there is a connected communication/cooperation network between the systems in the network, and the reference trajectory is available in a fraction of the locations, \textit{i. e.,} pinning locations. Finally, we consider a network of Lorenz oscillators with parameter mismatches to numerically verify our analytical results.

\section{Preliminaries}
%%%%%%%%%%%%%%%%%%%%%%%%%%%%%%%%
%%%%%%-----------------NOTATIONS
%%%%%%%%%%%%%%%%%%%%%%%%%%%%%%%%
\subsection{Notations and Background}
The set of real $n$-vectors is denoted by $\bbR^{n}$ and the set of real $m\times n$ matrices is denoted by $\bbR^{m\times n}$. We refer to the set of non-negative real numbers by $\bbR_{+}$. Matrices and vectors are denoted by capital and lower-case bold letters, respectively. Identity matrix is shown by \bI. The Euclidean ($\mathcal{L}_{2}$) vector norm is represented by $\lVert\cdot\rVert $.  Symmetric part of matrix, \bA, is denoted as $\bA^{(s)}\triangleq (\bA+\bA^T)/2$.

%------ Note
	 As it is known, the mismatched network (in general, mismatched systems), without compensators, cannot be absolutely synchronized; hence, the synchronization for these networks reduces to neighborhood synchronization, where the network trajectories will converge to a certain vicinity of each other and continue to stay there \cite{Sun09, Sorrentino11, Acharyya12, Manaffam.IET16}. To analyze this type of synchronization, the objective is to find the center and the radius of that neighborhood. In \cite{ Manaffam.IET16}, it has been shown that this center for \textit{undirected} networks is simple average of all the trajectories. This has also been used in \cite{Sun09, Sorrentino11, Acharyya12}. Consequently, the error of system $i$ from the average trajectory, $\bar{\bx} \triangleq \sum_{i=1}^{N}\bx_{i}/N$, yields, $\be_i\triangleq\sum_{j=1}^N(\bx_i-\bx_j ) / N$. Therefore, we can formalize the definition of $\varepsilon$-synchronization as follow.
	%------ Definition: e-synchronization
\begin{definition}[$\varepsilon$-synchronization] 
Let $\bx \triangleq [\bx_1^T \, \cdots \, \bx_N^T]^T$ denote the state of the network, where $N$ is the number of systems in the network. Then, the undirected network is $\varepsilon-$synchronized, iff
	 \begin{equation} \label{def: e-sync}
	 	\frac1N\lim_{ t \to \infty} \|(\bR_N \otimes \bI_n) \, \bx \| = \varepsilon,
	\end{equation}
	 where 
	 \begin{equation} \label{eq: Globally Connected  Net}
				\bR_N\triangleq\left[\begin{array}{ccccc}
							N-1 &-1&-1&\cdots&-1\\
							-1 & N-1&-1&\cdots&-1\\
							\vdots&\ddots&\cdots& &\vdots\\
							-1& -1&\cdots &-1 & N-1
							\end{array}\right]_{N\times N}.
	\end{equation}
\end{definition}
Please note that asymptotic absolute-synchronization is achieved if \cite{Restrepo04, Sun09, Sorrentino11, Acharyya12, Manaffam.IET16}\[\lim_{t\to\infty}\|\bx-\bar{\bx}\| =0.\]

	Here are several lemma's, which will be used later in the analysis.
\begin{lemma}\label{lem: commute} Any $N \times N$ symmetric Laplacian, \bP, and $\bR_N$ commute and moreover,
	\begin{equation}
		\bR_N \bP = \bP \bR_N = N \bP.
	\end{equation}
	\begin{IEEEproof}
		See Appendix \ref{proof: commute}.
	\end{IEEEproof}
\end{lemma}

\begin{lemma}\label{Lemma: Vector Product Inequality}
		Let \bx~and \by~to be any arbitrary vectors and \bK~to be a positive definite matrix and \bP~a matrix of proper dimensions. Then
			\begin{align*}
				\bx^T\bP\by+\by^T\bP^T\bx=2\bx^T\bP\by&\le\bx^T\bP\bK^{-1}\bP^T\bx+\by^T\bK\by.
			\end{align*}
\end{lemma}
\begin{lemma}\cite{Prasolov}\label{Lemma: Joint Diagonalization}
			If \bM~and \bK~commute, \textit{i. e.}, $\bM\bK=\bK\bM$, then they can be jointly diagonalized by a unitary matrix, \bQ~such that
			\begin{align*}
				\bM&=\bQ\bJ_M\bQ^T,\\
				\bK&=\bQ\bJ_K\bQ^T
			\end{align*}
			where superscript $T$ denotes Hermitian transpose. The diagonal entries of $\bJ_M$ and $\bJ_K$ are eigenvalues of $\bM$ and \bK, respectively.
\end{lemma}
	
\begin{lemma}\cite[{Theorem 4. 8}]{Khalil02}\label{Lemma: asymptotic W}
		 Suppose that $\bff(\bx)$ is continuous and satisfies \eqref{eq: boundF} and it is uniform in $t$. Let $V:\bbR^m\to\bbR$ be continuously differentiable function and continuous function $W(\bx)$ such that
		 \begin{align}
		 k_1\|\bx\|^{c_1}\le V(\bx)\le k_2\|\bx\|^{c_2}\\
		 \dot{V}(\bx)=\frac{\partial V}{\partial \bx}\bff(\bx)\le-W(\bx)\le 0
		 \end{align}
		 where $k_i$ and $c_i$ are positive constants. Then all solutions of 
		 \begin{align*}
			 &\dbx=\bff(\bx)\\
			 &\bx(t_0)=\bx^0,
		 \end{align*}
		 are ultimately bounded and 
		 \[\lim_{t\to\infty}W(\bx)=0.\]
	 \end{lemma}
	\begin{lemma}\cite[{Theorem 8. 2}]{Khalil02}\label{Lemma: Bounded x}
		 Suppose that $\bff(\bx)$ is continuous and satisfies \eqref{eq: boundF} and it is uniform in $t$. Let $V:\bbR^m\to\bbR$ be continuously differentiable function such that
		 \begin{align}\begin{array}{ll}
		 k_1\|\bx\|^{c_1}\le V(\bx)\le k_2\|\bx\|^{c_2}& \\
		 \dot{V}(\bx)=\frac{\partial V}{\partial \bx}\bff(\bx)\le-k_3\|\bx\|^{c_3} &\forall \|\bx\|\ge r\end{array}
		 \end{align}
		 where $k_i$ and $c_i$ are positive constants. Then there exists $t_1>t_0$ such that 
		 \begin{align*}\begin{array}{ll}
			 \|\bx\|\le k_4  \|\bx_0\| \exp(-c_4(t-t_0)),&\forall t_0\le t\le t_1\\
			\|\bx\|\le \left(\frac{k_2}{k_1}\right)^{1/c_1}r^{c_2/c_1}&\forall t>t_1.\end{array}
		 \end{align*}
	 \end{lemma}

%%%%%%%%%%%%%%%%%%%%%%%%%%%%%%%%
%%%%%%-----------------SYSTEM MODEL
%%%%%%%%%%%%%%%%%%%%%%%%%%%%%%%%

\subsection{Systems Model}
	Let the dynamics of networked systems be given as $ \forall i$
		\begin{align}\label{eq: NetworkEq1}
			&\dbx_i=\bff(\bx_i)+\bG(\bx_i)\bgamma_i+\sum_{j=1}^N a_{ij}\bH(\bx_j-\bx_i)+\bu_i\\
			&\bx_i(t_0)=\bx_i^0. \nonumber
		\end{align}
		where $\bx_i\in\Omega$ is the state vector of the system $i$, $\bx_i^0$ is the initial state of the system $i$, 
$\bff:\Omega \to \bbR^n$ describes the dynamics of the nominal system, and $\bu_i\in \bbR^n$ is the input vector. $\bG(\bx_i) \bgamma_i$ represents the uncertainty in the dynamics of system $i$. More precisely, $\bgamma_i\in \cP$ is the uncertainty/mismatch vector corresponding to the system $i$ and it is limited to the set $\cP$ with dimension $|\cP|=m$, where the uncertainties affect the individual systems according to the function $\bG:\bbR^n\to\bbR^{n\times m}$.  The adjacency matrix of the network is denoted by $\bA=[a_{ij}]$, where $a_{ij}\in\bbR$ indicates the weight of the connection from node $j$ to node $i$. There is no connection if $a_{ij}=0$. The term $a_{ij} \bH(\bx_i-\bx_j)$ indicates that the system $i$ is coupled to the system $j$, where $\bH \in \bbR^{n \times n}$~is the inner coupling matrix.	

Define the Laplacian/gradient matrix of the network, $\bL=[l_{ij}]$, as
	\begin{align}
		l_{ij}=\left\{\begin{array}{ll}
						-a_{ij}&i\ne j,\\
						\sum_{j=1}^Na_{ij}&i=j.
					 \end{array}\right.\label{eq: Laplacian}
	\end{align}
	\bL~is a zero-sum-row matrix and it is positive semidefinite. From this point on, we will represent the network by \bL. With this definition, \eqref{eq: NetworkEq1} can be rewritten as
	\begin{align}\label{eq: NetworkEq}
			&\dbx_i  = \bff(\bx_i) + \bG(\bx_i) \bgamma_i - \sum_{j=1}^N l_{ij}\bH\bx_j +\bu_i,
			\\
			&\bx_i(t_0)  =  \bx_i^0. 
			\nonumber
	\end{align}
In the remainder of the paper, we will assume that the followings hold.
	\begin{assumption}\label{Assumption: Connected}
	The plant network represented by the Laplacian matrix, \bL,~is connected and undirected.
	\end{assumption}
	The connectivity of the network implies that the Laplacian matrix in \eqref{eq: Laplacian}, has only one zero eigenvalue \cite{Mohar91}. The network being undirected implies that the Laplacian is symmetric, \textit{i .e.,} $\bL = \bL^T$. Therefore, all its eigenvalues, $\mu_{i}$, are non-negative real numbers. Please note that we do note require the weights of the connection to be binary, hence, we consider the general class of weighted-undirected networks.
	\begin{assumption}\label{Assumption: BoundF}
		There exists a positive semidefinite matrix \bF~such that following inequality holds 
			\begin{align}
				(\tilde{\bx}-\tilde{\bs})^T[\bff(\tilde{\bx})-\bff(\tilde{\bs})]\le(\tilde{\bx}-\tilde{\bs})^T\bF(\tilde{\bx}-\tilde{\bs}),\label{eq: boundF}
			\end{align}
			for all $(\tilde{\bx},\tilde{\bs})\in\Omega\times\Omega$.
	\end{assumption}	
	Note that this assumption is not very restrictive: if all the elements of the Jacobian of $\bff(\bx)$ with respect to state vector, \bx, is bounded, there always exists a positive semidefinite matrix \bF~such that assumption \eqref{eq: boundF} holds \cite{Yu09}. As discussed in \cite{DeLellis11}, this assumption is closely related to QUAD$-$condition. 		

	\begin{assumption}[Bounded uncertainties]\label{Assumption: Bounded Mismatch}
		If $(\bx,\, \bgamma)\in \Omega \times \cP$, there exists a symmetric positive semidefinite matrix \bGamma~and vector $\bgamma_c$ such that following inequality holds 
			\begin{align}
				\bgamma^T \bG(\bx)^T \bG(\bx) \bgamma \le \bgamma_c^T \bGamma \gamma_c,\quad \forall (\bx ,\, \bgamma)\in \Omega \times \cP.
			\end{align}
	\end{assumption}	
This assumption basically states that the uncertainties, $\bgamma_i$'s, are bounded.
%*******************************************%		
%------------------------Problem Statement	
%*******************************************%	
\subsection{Problems Statements}
As discussed, since there are uncertainties in the network, in general, the network does not achieve absolute synchronization. Thus, based on the described system model, there are two natural questions to be asked:
\textit{1) what are the conditions that the network in \eqref{eq: NetworkEq} should satisfy to achieve $\varepsilon$-synchronization? And if it $\varepsilon$-synchronizes, what is the bound on the norm total error from the average trajectory, $\varepsilon$? 2) How does the network in \eqref{eq: NetworkEq} can be pinned such that it asymptotically converges to a known reference trajectory?} 

Next, we formalize these questions as
\begin{problem}
a) If Assumptions  \ref{Assumption: Connected}-\ref{Assumption: Bounded Mismatch} hold. Find conditions on $\mathbf{L}$ such that there exists a positive constant, $\varepsilon > 0$, that in finite time, $ t_\varepsilon > t_0$, we have
	\begin{align}
		\frac1N \|(\mathbf{R}_N \otimes \mathbf{I}_n) \, \mathbf{x}\| \le \varepsilon \quad \forall t  >  t_\varepsilon  >  t_0;
	\end{align}
b) If $\varepsilon-$synchronization occurs, what is $\varepsilon$?
\end{problem}

\begin{problem}
	For a given reference trajectory, \bs, 
		\begin{align}\label{eq: reference}
			&\dbs=\bff(\bs),\\
			&\bs(t_0)=\bs^0,\nonumber
		\end{align}
	find a control law, $\bu_i$, such that the network in \eqref{eq: NetworkEq} asymptotically converges to \bs, that is,
		\begin{equation}
			\lim_{t \to \infty} \|\bx - \mathbf{1}_N \otimes \mathbf{s}\| = 0.
		\end{equation}
\end{problem}

%*******************************************%		
%------------------------Main Results
%*******************************************%		
\section{Analytical Results}
In this section, first we derive the sufficient conditions on bounded stability networked systems with uncertainties. Then, using decentralized control and assumption of constant uncertainties, the mismatched parameters are compensated and the network is driven to the reference trajectory.
\subsection{Boundedness of The Synchronization Error}
	In this section, we will show that the error of the network in \eqref{eq: NetworkEq} from its average trajectory, $\bar{\bx} = \sum\limits_{i = 1}^N \bx_i/N$, is ultimately bounded. Additionally, we will derive an upper bound on the norm of that error.
\subsubsection{Error development}
		let us define the synchronization error as $\be_i \triangleq \bx_i - \bar{\bx}$. The error dynamics for $\forall i$ can be expressed as
		\begin{align}
			\dbe_i=&\frac1N(\bR_N \otimes \bI_n) \, \dbx \nonumber\\
			=& \frac1N\sum_{j=1}^N\left[\Big(\bff(\bx_i)-\bff(\bx_j)\Big)+\Big(\bG(\bx_i)\bgamma_i - \bG(\bx_j) \bgamma_j\Big)\right]\nonumber\\
			& - \sum_{j =1}^N l_{ij}\, \bH \,\bx_j + \frac1N\sum^N_{i,j =1} l_{ij}\, \bH \,\bx_j \nonumber \\
			=& \frac1N\sum_{j=1}^N\left[\Big(\bff(\bx_i)-\bff(\bx_j)\Big)+\Big(\bG(\bx_i)\bgamma_i - \bG(\bx_j) \bgamma_j\Big)\right]\nonumber\\
			& - \sum_{j =1} l_{ij}\, \bH \,\bx_j  .\label{eq: NetworkError}
		\end{align}
		Please note that from Assumption \ref{Assumption: Connected} the Laplacian of the network is symmetric, and henceforth zero column-sum, thus, we have $\sum^N_{i,j =1} l_{ij}\, \bH \,\bx_j = \b0$, and the last equality follows.
		
\subsubsection{Results and Discussion}	
	\begin{theorem}\label{Theorem: BoundedError}
	 	Let Assumptions \ref{Assumption: Connected}-\ref{Assumption: Bounded Mismatch} hold, if there exists a positive constant, $\lambda$ such that
		\begin{align}
			\bF-{\mu_i}\bH^{(s)}+\lambda\bI_{n}\prec\b0,\quad \forall i = 1, \,\cdots,\,N-1\label{eq: ConditionTheorem1}
		\end{align}
	 	where $\mu_i$'s are eigenvalues of the Laplacian of the network sorted descendingly, then the error in \eqref{eq: NetworkEq} is ultimately uniformly bounded around $\bar{\bx} = \sum_{i =1}^N \bx_i/N$. Furthermore, the synchronization error in \eqref{eq: NetworkError}, 				$\be  \triangleq [ \be_1^T \, \cdots \, \be^T_N ]^T$, is bounded as
	 	\begin{align}\label{eq: Theorem1}
		 	\|\be\|  \le & \quad \sqrt{ {N}{\bgamma_c^T\bGamma\bgamma_c}/{\lambda^\star}^2},\\
			\lambda^{\star}=&~\max \quad\lambda  \label{eq: lambda_star}\\
				~&~\text{s. t.} \quad \bF-{\mu_i}\bH^{(s)}+\lambda\bI_n\prec\b0,\quad \forall i = 1, \,\cdots,\,N-1\nonumber
		 \end{align}
		 in finite time.
	 	
		\begin{IEEEproof} 
			See appendix \ref{Proof: BoundedError}.
		\end{IEEEproof}
	\end{theorem}
	\begin{remark}
		The existence of positive $\lambda>0$ that satisfies \eqref{eq: ConditionTheorem1} is a sufficient condition on existence and convergence to the bound in \eqref{eq: Theorem1}. Additionally, it should be noted that according to Lemma \ref{Lemma: Bounded x}, the network reaches the bound \eqref{eq: Theorem1} in finite time.
	\end{remark}
	\begin{remark}
	Since Assumption \ref{Assumption: Bounded Mismatch} does not require the uncertainties to be constant, the results in Theorem \ref{Theorem: BoundedError} hold if the uncertainties are time varying but bounded.
	\end{remark}
		In contrast to the previous results reported in \cite{Restrepo04, Sun09, Sorrentino11, Acharyya12, Manaffam.IET16}, Theorem \ref{Theorem: BoundedError} guarantees convergence in finite time. Although conservative, the conditions in \eqref{eq: ConditionTheorem1} does not require linearization around average trajectory which leads to calculation of transition matrix of the network error or Lyapunov exponents. The same is true for the bound on the error in \eqref{eq: Theorem1}. Consequently, this theorem renders the stability analysis of the networked systems much simpler and straightforward. 
%\begin{corollary}[$\varepsilon-$Synchronization in Erd\"os-R\'enyi Networks] Let Assumptions \ref{Assumption: Connected}-\ref{Assumption: Bounded Mismatch} hold,
%		
%\end{corollary}
		
\subsection{Compensation for constant uncertainties}
	In this part, first, we employ decentralized control to stabilize the network and compensate for the constant uncertainties in the network.
	\subsubsection{Error development}	
		if the dynamics of the reference trajectory is given by \eqref{eq: reference}, then the error dynamics from the reference, \bs, can be expressed as $\forall i = 1, \cdots, N$
		\begin{align}\label{eq: Error_Ref}
			\dbe_i = \bff(\bx_i) - \bff(\bs) + \bG(\bx_i) \bgamma_i -  \sum_{j =1}^N l_{ij} \bH\,\bx_j + \bu_i.
		\end{align}
	\subsubsection{Results and discussion}	
		In the rest of the paper, we will assume that all the nodes have compensators and all the uncertainties are constant. The results for asymptotic convergence of the synchronization error from the reference signal will be presented in two fold: decentralized and distributed.
		
		\begin{theorem}[Decentralized Compensation]\label{Theorem: MismatchEstimation}
		Let Assumptions \ref{Assumption: Connected} and \ref{Assumption: BoundF} hold and the uncertainty vectors, $\bgamma_i$'s, be constant. Then, the network in \eqref{eq: NetworkEq} with $\forall i = 1,\,\cdots,\, N$
		\begin{align}\label{eq: input}
			\bu_i&=-z_i\bH(\bx_i-\bs)-\bG(\bx_i)\hat{\bgamma}_i,\\
		\label{eq: Estimation}
			\dot{\hat{\bgamma}}_i&=k_i\bG^T(\bx_i)(\bx_i-\bs),
		\end{align}
		 asymptotically uniformly converges to the reference signal, \bs, if there exists positive constants, $k_i > 0$, and a diagonal matrix, $\bZ=\mbox{diag}([z_1,\,\cdots,\,z_N])$, $\forall z_i>0$ , such that
				\begin{align}\label{eq: Theorem2}
					\bF-\mu_i\bH^{(s)}\prec\b0 ,\quad i = 1,\, \cdots, \, N
				\end{align}
				where $\mu_i$'s are eigenvalues of $\bL+\bZ$ and $\bZ \triangleq \diag([z_1 \, \cdots \, z_N]^T)$.\\
		\textbf{Proof:} See appendix \ref{Proof: MismatchEstimation}.
	\end{theorem}
	The first term in \eqref{eq: input} is a common feedback control used in pinning control of identical networked systems\cite{Chen07, Manaffam.TCAS13, Manaffam.ACC13,Yu09}. The second term in conjunction with \eqref{eq: Estimation} estimates and compensates for the parameter mismatches of the systems. The detail of choosing the estimator in \eqref{eq: Estimation}, is given in Appendix \ref{Proof: MismatchEstimation}.
	\begin{remark}\label{remark: Conv_Est}
		From Theorem \ref{Theorem: MismatchEstimation}, we have
\[			\lim_{t \to \infty} [ \dbx_i -  \bff(\bx_i)] = \dbs - \bff(\bs) = \b0,\]
and
\[			\lim_{t \to \infty} \bG (\bx_i)  = \bG(\bs).\]
		Now, if there exists $T \ge t_0$ such that $\bG(\bs)$ is not singular for $t > T$, then from \eqref{eq: NetworkEq} we can conclude that \[\lim_{t \to \infty} \hat{\bgamma_i} = \bgamma_i.\]
	\end{remark}
			
As it is clear from Theorem \ref{Theorem: MismatchEstimation}, all the nodes are required to have the reference trajectory \eqref{eq: reference}, however, in spatially distributed networks to provide the reference trajectory to all the systems is costly and impractical. To alleviate this issue, it is more convenient to have the systems receive and send information to their neighboring systems and provide the reference trajectory to small fraction of the network. The idea is to \textit{cooperatively} estimate the mismatches. Next theorem is devised to address this drawback of Theorem \ref{Theorem: MismatchEstimation}.

\begin{theorem}[Distributed Compensation]\label{Theorem: Distributed Estimation}
Let Assumptions \ref{Assumption: Connected} and \ref{Assumption: BoundF} hold and the uncertainty vectors, $\bgamma_i$'s, be constant. Let $\bB = [b_{ij}]$ and $\bC = [c_{ij}]$ be the Laplacian matrices corresponding of two graphs on $1,\, \cdots,\,N$, where $\bC$ is connected and undirected and $ \forall i, \, j: c_{ij} \le 0$, if the input of the system $i$ is selected as
\begin{eqnarray}
	\bu_i &= &-\sum_{j =1}^N b_{ij} \bH \bx_j - g_i \bH (\bx_i - \bs) - \bG(\bx_i)\hat{\bgamma}_i \label{eq: Distributed_input}\\
	\dot{\hat{\bgamma}}_i &= &k_i \bG^T(\bx_i) \left(\sum_{j = 1}^N c_{ij} \bx_j \,+\, z^\prime_i (\bx_i - \bs) \right),\label{eq: Distributed_Estimation}
\end{eqnarray}
 then, the network in \eqref{eq: NetworkEq} asymptotically uniformly converges to the reference signal, if there exists positive constants, $k_i>0$, and nonnegative constants $z_i,z_i^\prime \ge0$ such that at least one $z^\prime_i >0$ and $z_i >0$ and 
				\begin{align}\label{eq: Theorem3}
					\bF- \mu_i \bH^{(s)}\prec\b0, \quad i = 1\,\cdots, \, N
				\end{align}
				where $\mu_i$'s are eigenvalues of $\bL+\bB+\bZ$ and $\bZ \triangleq \diag([z_1 \, \cdots \, z_N]^T)$.\\
				\textbf{Proof:} See Appendix \ref{proof: 3}.
\end{theorem}
	
		\begin{remark}
		It should be noted that the connectivity of the Laplacian $\bB$ is not required in Theorem \ref{Theorem: Distributed Estimation}, in fact, in sufficiently connected/coupled networks the condition \eqref{eq: Theorem3} can be satisfied with $\bB = \b0$, as will be shown in the numerical example. In this case, the problem of synchronization of the network to the reference trajectory, is reduced to the well-known pinning problem. However, Theorem \ref{Theorem: Distributed Estimation} requires the communication/feedback network of mismatch estimation, \bC, in \eqref{eq: Distributed_Estimation} to be connected. Although this condition is related to the validity of the chosen Lyapunov condition, intuitively, it also seems required as to estimate the mismatches, each system should have some level of information about the reference system which can collect from its neighbors on the feedback network, \bC, and  if the network is not connected this might not be satisfied. Furthermore, Theorem \ref{Theorem: Distributed Estimation} requires $ \forall i, \, j: c_{ij}\le 0$ which implies that the feedback from the neighboring systems should be positive while there is an equivalent negative self-feedback. This is standard condition in cooperative control. 
		\end{remark}
		\begin{remark}For practical purposes, if condition \eqref{eq: Theorem3} allows, the communication network for the linear part of the controller corresponding to the Laplacian $\bB$ can be assumed to be a subnetwork of the communication network of distributed mismatch estimators corresponding to $\bC$. Also in application such as microgrid for distributed generators, due to availability of communication protocols such as power line communication (PLC), it is reasonable to assume that the communication network corresponding to \bC, is a subnetwork of plant network corresponding to \bL.
	\end{remark}
	%Theorem \ref{Theorem: Distributed Estimation} states that as the mismatched parameters of each isolated system is compensated by \eqref{eq: Distributed_Estimation}, 
	\begin{remark}
		The condition of $\bZ\succeq \b0 $ and $\bZ \ne \b0$, is a necessary condition in pinning problems and there are many literatures dealing with the design of $\bZ$, please see \cite{Chen07}\cite{Yu09}\cite{ Manaffam.ACC13}\cite{Pirani.TAC16}\cite{ Bapat.Cyb16}\cite{ DeLellis13} and references therein. The same applies to $\bZ^\prime$.  
	\end{remark}
	
		\begin{remark}\label{remark: Conv_Est2}
			Similar to Remark \ref{remark: Conv_Est}, if there exists a $T>t_0$ such that  $\bG(\bs)$ is not singular, \textit{i. e.,} $\text{det}(\bG(\bs))\ne0:~ \forall t \ge T$, then $\lim_{t \to \infty} \hat{\bgamma}_i =  \bgamma_i$.
		\end{remark}

\section{Numerical Example}
Consider the Lorenz system
\begin{align*}
	\dbx=\left[\begin{array}{c}
						a(x_2-x_1)\\
						bx_1-x_2-x_1x_3\\
						x_1x_2-cx_3
				 \end{array}\right]+\left[\begin{array}{ccc}
				 						x_2-x_1  & 0& 0\\
				 						0 & x_1 & 0\\
				 						0 & 0& -x_3
				 				 \end{array}\right]\bgamma,
	\end{align*}
	where $\bgamma=[\gamma_1~\gamma_2~\gamma_3]^T$, $(a,b,c)=(10,\,28,\,8/3)$. We assume a network of size $N = 50$ which is globally connected, \textit{i. e.,} $\bL=\bR_{50}$, where $\bR_N$ is defined in \eqref{eq: Globally Connected  Net}. The inner coupling matrix is assumed to be $\bH=  10 \, \bI_3$. The mismatch parameters are assumed to be constant and satisfy $|\gamma_{1,i}|\le 0.1\,a$, $|\gamma_{2,i}|\le 0.1\,b$, $|\gamma_{3,i}|\le 0.1\,c$. 

	The matrices \bF~and \bGamma, in Assumptions \ref{Assumption: BoundF} and \ref{Assumption: Bounded Mismatch} are \cite{Manaffam.Allerton15}
	\begin{align*}
	\bF=&\left[\begin{array}{ccc}
					21 & 10 & 0   \\
					28 & 23 & 0   \\
					0   & 0   & 40
			   \end{array}\right],    
	\bGamma=&\left[\begin{array}{ccc}
						213  & 0& 0\\
						0 & 400 & 0\\
						0 & 0& 2500
				  \end{array}\right].
	\end{align*}
\begin{figure}
	\includegraphics[width =8.9 cm]{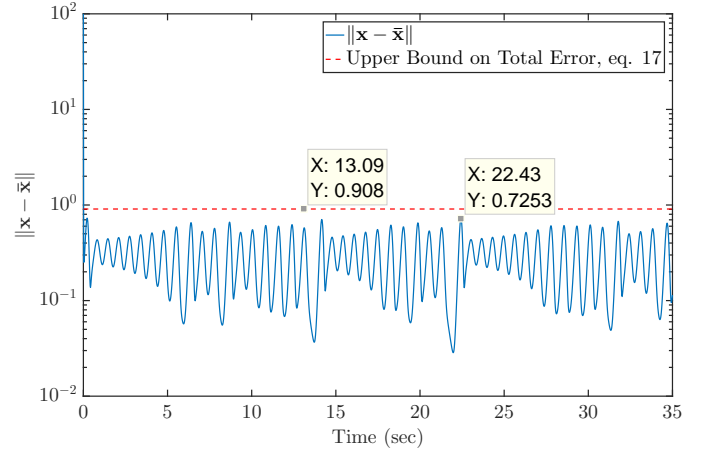}
	\caption{The evolution of the networked systems' error from its average trajectory, and the bound given in \eqref{eq: Theorem1}.}\label{fig: Bounded_Error}
\end{figure}	
Fig. \ref{fig: Bounded_Error} shows the error of the networked system from its average trajectory. As it can be seen the error from the average trajectory reaches the bound in \eqref{eq: Theorem1} in finite time as expected from Lemma \ref{Lemma: Bounded x} and Theorem \ref{Theorem: BoundedError}. The solid plot corresponds to norm of total error, \textit{i. e.,} $\|\be\|$, as defined in Theorem \ref{Theorem: BoundedError}, and dashed plots corresponds to several samples of system trajectory errors, $\|\be_i\|$. The bound in \eqref{eq: Theorem1} is $0.91$.

\begin{figure}[t]
	\includegraphics[width = 8.9cm]{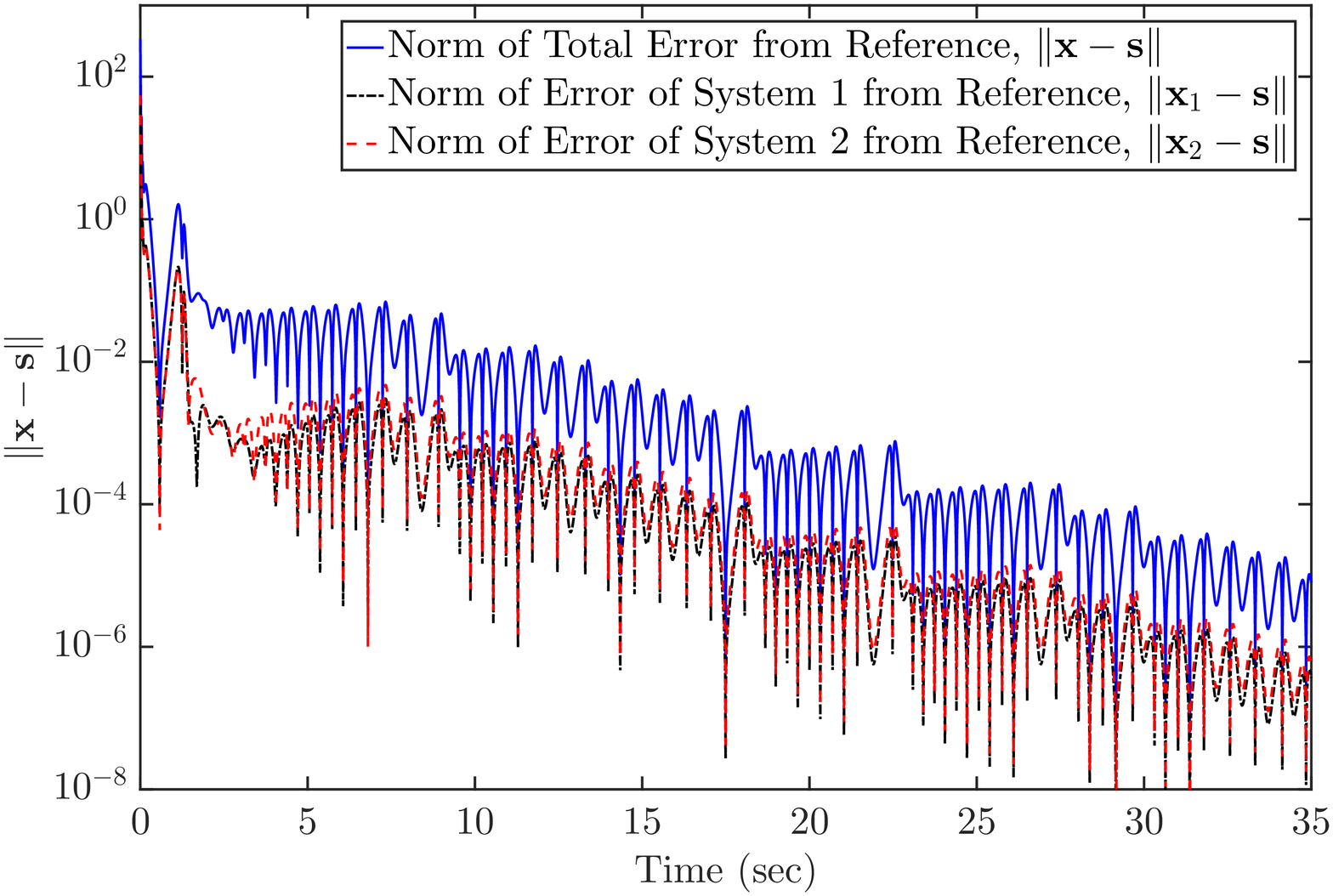}	\caption{The evolution of norm of network 's error from the reference trajectory, $\| \be\| = \|\bx - \bone_N\otimes \bs\|$, and sample error trajectories for systems 1 and 2, \textit{i. e.,} $\|\be_1\| = \|\bx_1 - \bs\| $ and $ \|\be_2 \| = \| \bx_2 - \bs\|$ using decentralized controller \eqref{eq: Theorem2} in Theorem \ref{Theorem: MismatchEstimation}.}\label{fig: Decentral_Esti_States}\vspace{+ 0.5 cm}
	\includegraphics[width = 8.9cm]{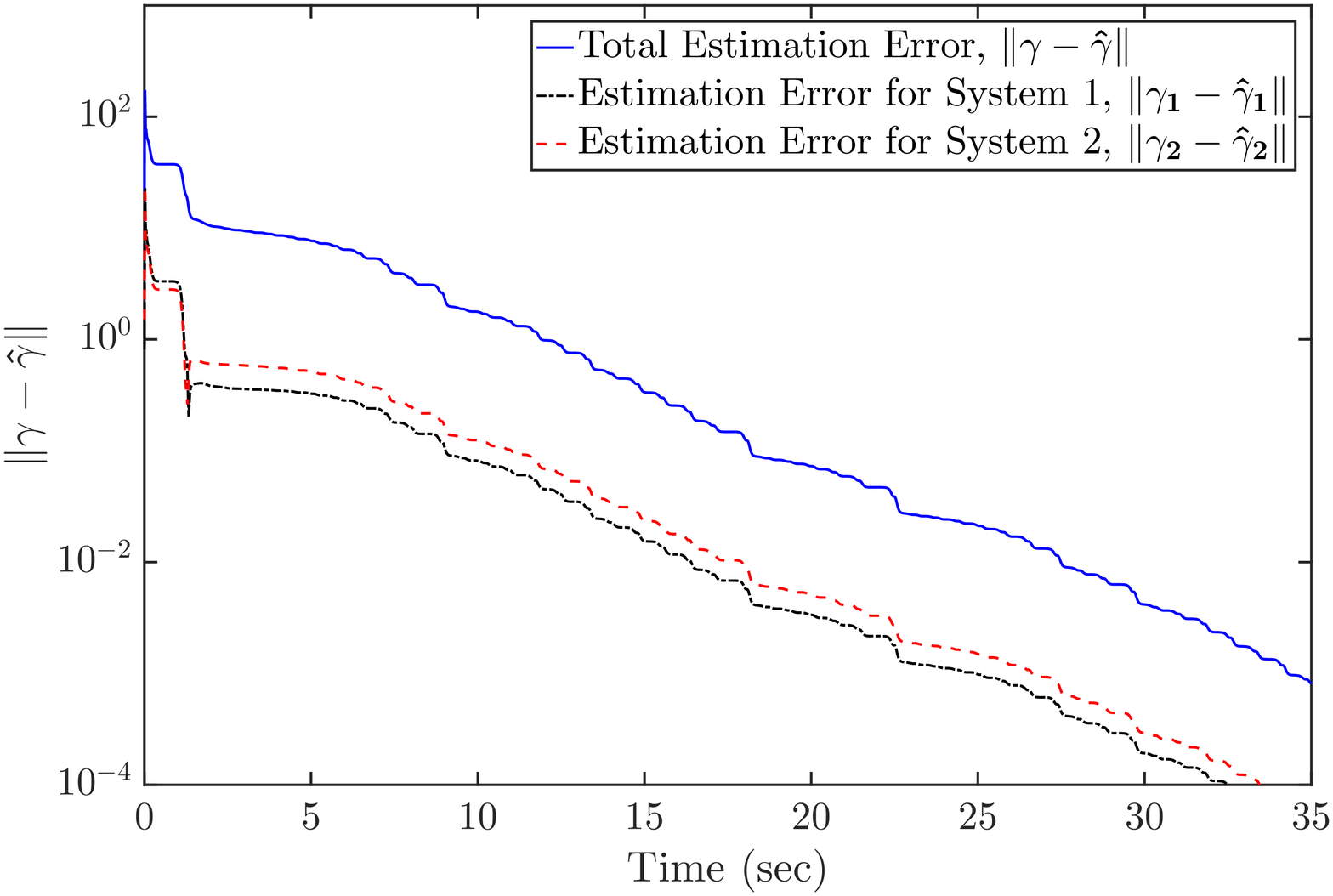}\caption{The evolution of norm of network 's estimation error from the mismatches, $\| \tilde{\bgamma}\| = \|\hat{\bgamma} - \bgamma\|$, and sample estimation errors for systems 1 and 2, \textit{i. e.,} $\|\tilde{\bgamma}_1\| = \|\hat{\bgamma}_1 - \bgamma_1\| $ and $ \|\tilde{\bgamma}_2 \| = \| \hat{\bgamma}_2 - \bgamma_2\|$ using decentralized estimator \eqref{eq: input} in Theorem \ref{Theorem: MismatchEstimation}.}\label{fig: Decentral_Esti_Parameters}
\end{figure}

The norm of network error, $\|\be\|$ and norm of total error for estimation, $\|\hat{\bgamma}-\bgamma\|$, are shown in fig.s \ref{fig: Decentral_Esti_States} and \ref{fig: Decentral_Esti_Parameters}, respectively, where the compensator \eqref{eq: Theorem2} in Theorem \ref{Theorem: MismatchEstimation} is used on all the locations in the network. The controller parameters are set as follow: $ z_i = 10,\, k_i =1, \,\forall \,i $. The dashed plots provide several examples of the norm of synchronization error, $\|\be_i\|$, and estimation error, $\|\hat{\bgamma}_i-\bgamma_i\|$, at each system, respectively. As it can be seen and predicted by Theorem \ref{Theorem: MismatchEstimation}, the synchronization error from the reference trajectory asymptotically vanishes. Moreover, since $\bG(\bs)$ is nonsingular, all the estimated uncertainties converge to their actual values, this has been predicted in Remark \ref{remark: Conv_Est} and shown in fig. \ref{fig: Decentral_Esti_Parameters}.
	
\begin{figure}[t]
	\includegraphics[width =8.9cm]{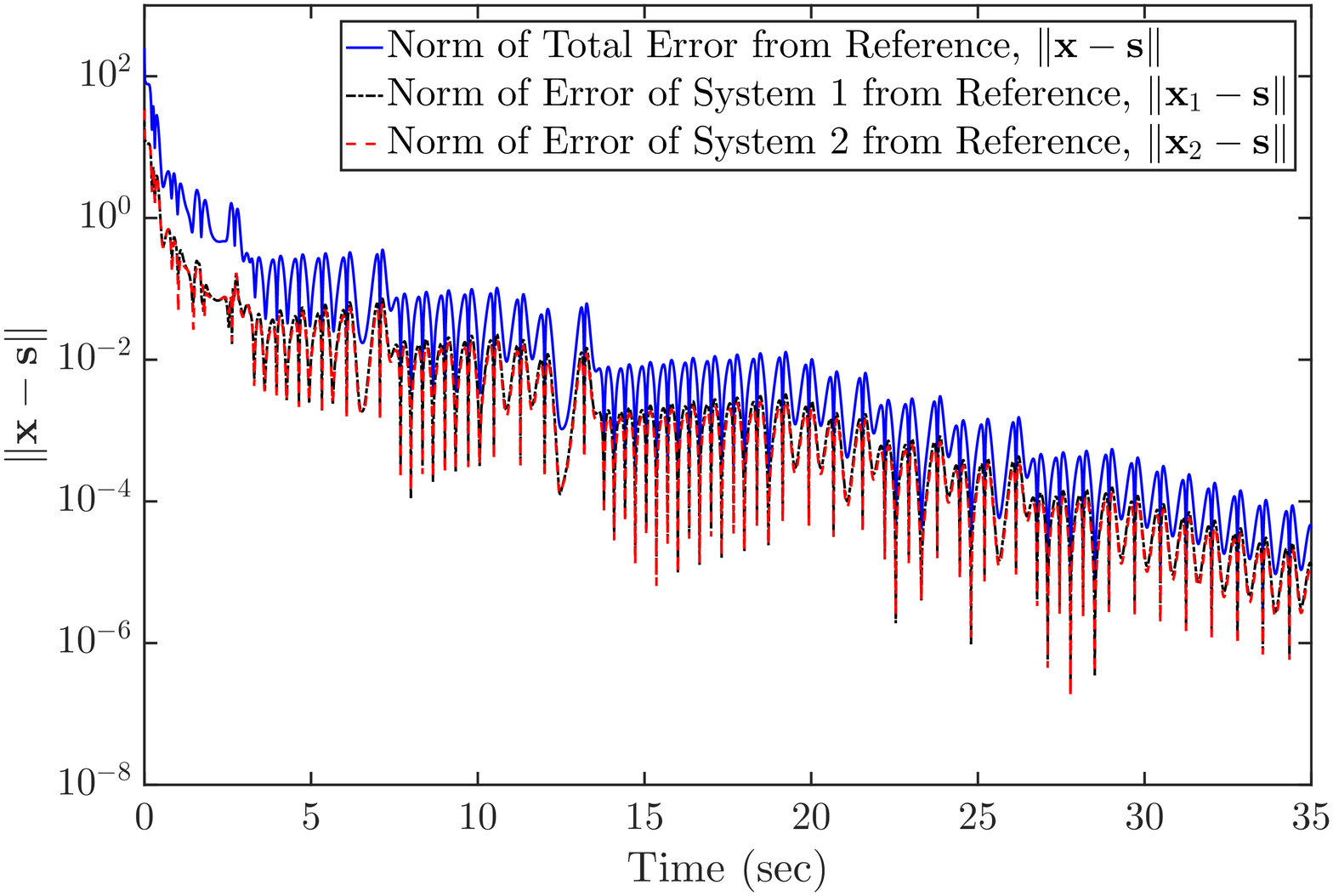}
		\caption{The evolution of norm of network 's error from the reference trajectory, $\| \be\| = \|\bx - \bone_N\otimes \bs\|$, and sample error trajectories for systems 1 and 2, \textit{i. e.,} $\|\be_1\| = \|\bx_1 - \bs\| $ and $ \|\be_2 \| = \| \bx_2 - \bs\|$ using distributed controllers \eqref{eq: Distributed_input} in Theorem \ref{Theorem: Distributed Estimation}.}\label{fig: Dist_Esti_States}\vspace{+ 0.5 cm}
	\includegraphics[width =8.9cm]{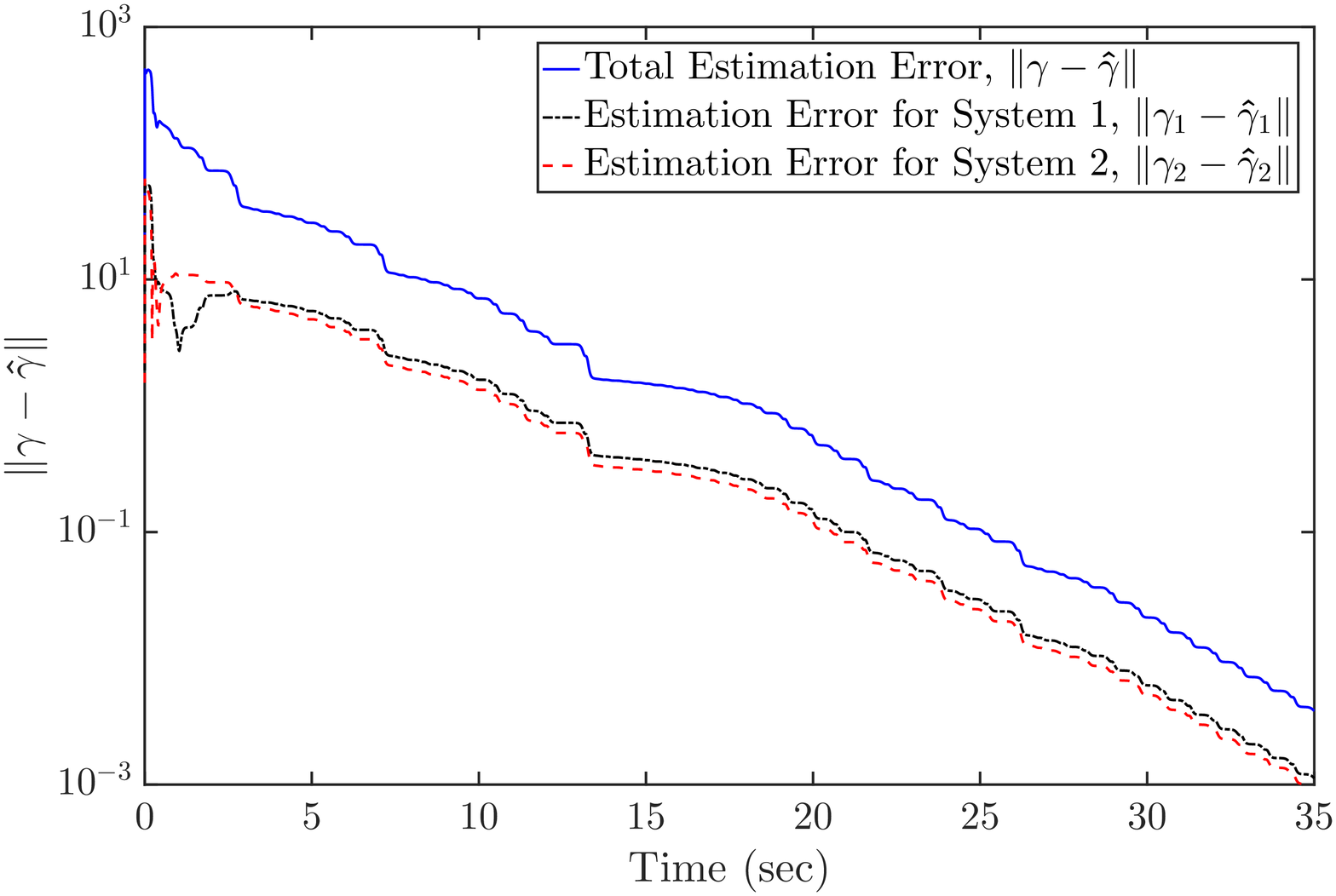}
	\caption{The evolution of norm of network 's estimation error from the mismatches, $\| \tilde{\bgamma}\| = \|\hat{\bgamma} - \bgamma\|$, and sample estimation errors for systems 1 and 2, \textit{i. e.,} $\|\tilde{\bgamma}_1\| = \|\hat{\bgamma}_1 - \bgamma_1\| $ and $ \|\tilde{\bgamma}_2 \| = \| \hat{\bgamma}_2 - \bgamma_2\|$ using distributed estimators \eqref{eq: Distributed_Estimation} in Theorem \ref{Theorem: Distributed Estimation}.}\label{fig: Dist_Esti_Parameters}
\end{figure}

Fig.s \ref{fig: Dist_Esti_States} and \ref{fig: Dist_Esti_Parameters} show the evolution of the norm of synchronization error, $\|\be\|$ and norm of total error for estimation, $\|\hat{\bgamma}-\bgamma\|$, respectively, where the compensator \eqref{eq: Theorem3} in Theorem \ref{Theorem: Distributed Estimation} is used. The communication network, $\bC = [c_{ij}]$, assumed to be a path graph, \textit{i. e.,} 
\begin{equation*}
c_{ij} = \left\{\begin{array}{ll}
			-1 & j = i +1\, \& \, i = 1, \,\cdots \, N - 1\\
			-1 & j = i -1\, \& \, i = 2, \,\cdots \, N\\
			2  & j = i , \& \, i = 2, \,\cdots \, N -1\\
			1 & j = i , \& \, i =  1, \, N\\
			0 & \text{o. w.}
	\end{array}\right.
\end{equation*}
Considering the calculated $\bF$ in Assumption \ref{Assumption: BoundF} and given $\bH = 10 \, \bI_n$, the condition \eqref{eq: Theorem3} in Theorem \ref{Theorem: Distributed Estimation} is satisfied iff $\mu_i \ge 4.0\, \forall\, i$. Here, we assume $\bB = \b0$, hence the stability condition $\mu_i \ge 4.0\, \forall\, i$ implies that the network at least should be pinned on $5$ locations \cite{Pirani.TAC16, Manaffam.CDC13}. To increase the convergence of the estimator in \eqref{eq: Distributed_Estimation}, we have chosen to optimally pin the communication network, \bC, which results in pinning the systems $i  = 5 ,\,   16 ,   \,26  , \, 35  , \, 46$, \textit{i. e.,} $z_i^\prime = z_i = 1,\,\forall \,i\in\{5 ,\,   16 ,   \,26  , \, 35  , \, 46\}$ and otherwise $z^\prime_i =z_i =0$. The estimator gains are set to $k_i =10, \forall \,i$.

The dashed plots give several examples of the norm of synchronization errors, $\|\be_i\|$, and estimation errors, $\|\hat{\bgamma}_i-\bgamma_i\|$, for each system, respectively. As it can be observed from fig. \ref{fig: Dist_Esti_States} and predicted by Theorem \ref{Theorem: Distributed Estimation}, the synchronization error from the reference trajectory asymptotically vanishes. Moreover, As predicted in Remark \ref{remark: Conv_Est2}, since $\bG(\bs)$ is nonsingular, all the estimation errors for mismatched parameters asymptotically go to zero, this has been shown in fig. \ref{fig: Dist_Esti_Parameters}.

\section{Conclusion}
{Here, we have derived sufficient conditions on the ultimately bounded stability of network of mismatched systems. A bound on the error from the average trajectory of the networked system has been calculated, which the network achieves that bound in finite time. To pin the network to a reference trajectory/state, two adaptive controllers, with decentralized and distributed structures, have been proposed. It has been shown that the compensated network will achieve absolute synchronization in presence of constant parameter uncertainties.}

\appendices
%%-------------------
\section{Proof of Lemma \ref{lem: commute}}\label{proof: commute}
	Let $\bY = [y_{ij}] = \bR_N \bP$, then for $ \forall i \ne j$
	\begin{eqnarray}
		y_{ij} & = & \sum_{ k =1}^N r_{ik} p_{kj} = r_{ii} p_{ij} + r_{ij} p_{jj} + \sum_{k =1,k \ne i,\,j}
^N r_{ik} p_{kj}\nonumber\\
			& = & (N - 1) p_{ij} - p_{jj} - \sum_{k =1,k \ne i,\,j}
^N p_{kj},\nonumber 
	\end{eqnarray}
	considering that \bP~is \textit{symmetric} Laplacian, and henceforth, zero column-sum: $- \sum_{k =1,k \ne i,\,j}^N p_{kj} = p_{jj} + p_{ij}$, we have
	\[ y_{ij} = N p_{ij}.\]
	For $ i = j$ 
	\begin{eqnarray}
		y_{ii} & = & \sum_{ k =1}^N r_{ik} p_{ki} = r_{ii} p_{ii}  + \sum_{k =1,k \ne i}
^N r_{ik} p_{ki}\nonumber\\
			& = & (N - 1) p_{ii} - \sum_{k =1,k \ne i}
^N p_{ki},\nonumber 
	\end{eqnarray}
	using the fact that for \textit{symmetric} Laplacian matrices, $p_{ii} = - \sum_{k =1,k \ne i,\,j}^N p_{ki}  $, we have
	\[ y_{ii} = N p_{ii},\]
	which means $\bR_N\bP = N \bP$. Showing that $\bP\bR_N =N\bP$ is very similar to the above, except that $- \sum_{k =1,k \ne i,\,j}^N p_{jk} = p_{jj} + p_{ij}$ which it is true for any Laplacian.
%%-------------------
\section{Proof of Theorem \ref{Theorem: BoundedError}}\label{Proof: BoundedError}
	\begin{IEEEproof}
		Let $\bu = \b0$, $\be \triangleq [\be_1^T \, \cdots \, \be_N^T]^T$ and the candidate Lyapunov function be $V(\be)=1/2\be^T\be=1/2\sum_{i=1}^N\be_i^T\be_i$, then,
		\begin{align*}
			\dot{V}=& 1/2\sum_{i=1}^N\be_i^T\dbe_i+1/2\sum_{i=1}^N\dbe_i^T\be_i									\\
				   =& \frac1{N^2}\sum_{i=1}^N\left(\sum_{j=1}^N(\bx_i-\bx_j)^T\sum_{k=1}^N[\bff(\bx_i)-\bff(\bx_k)]\right)   \nonumber   \\ 
				   	&-\sum_{i,j=1}^Nl_{ij}\be_i^T\bH^{(s)}\be_j+\sum_{i=1}^N\be_i^T\bG(\bx_i)\bgamma_i 							\nonumber   
		\\
						   =&\frac1{N^2}\sum_{i,j,k=1}^N(\bx_i-\bx_j)^T[\bff(\bx_i)-\bff(\bx_k)]\nonumber  \\
				   &-\sum_{i,j=1}^Nl_{ij}\be_i^T\bH^{(s)}\be_j+\sum_{i=1}^N\be_i^T\bG(\bx_i)\bgamma_i.
		\end{align*}
		Since the $\sum_{i,j=1}^{N}(\bx_{i}-\bx_{j})=\b0$, the first sum, referred to as $V_1$, can be rewritten as
		\begin{align*}
			V_1=&\frac1N\sum_{i,j=1}^N(\bx_i-\bx_j)^T\bff(\bx_i)	\\
			   =&\frac1{2N}\sum_{i,j=1}^N(\bx_i-\bx_j)^T\bff(\bx_i)-\frac 1{2N} \sum_{i,j=1}^N (\bx_j-\bx_i)^T \bff(\bx_j)																  						 \\
			   =&\frac1{2N}\sum_{i,j=1}^N(\bx_i-\bx_j)^T[\bff(\bx_i)-\bff(\bx_j)]	 							\\
			   \stackrel{(a)}{\le}&\frac1{2N}\sum_{i,j=1}^N(\bx_i-\bx_j)^T\bF(\bx_i-\bx_j)										\\
			   \le&\frac1{2N}\sum_{i,j=1}^N(\be_i-\be_j)^T\bF(\be_i-\be_j)=\frac1{N}\sum_{i,j=1}^N\be_i^T\bF\be_i-\be_i^T\bF\be_j																					 \\
			   \le&\frac1N\be^T(\bR_N\otimes\bF)\be.
		\end{align*}
		where inequality $(a)$ is due to Assumption \ref{Assumption: BoundF}. Using Lemma \ref{Lemma: Vector Product Inequality} and Assumption \ref{Assumption: Bounded Mismatch}, 
		\begin{align*}
			\sum_{i=1}^N\be_i^T\bG(\bx_i)\bgamma_i&\le \sum_{i=1}^N\frac{\beta}2\be_i^T\be_i+ \frac1{2\beta}\bgamma_i^T\bG(\bx_i)^T\bG(\bx_i)\bgamma_i			\\
			&\le\frac{\beta}2\be^T\be+\frac N{2\beta}\bgamma_c^T\bGamma\bgamma_c,
		\end{align*}
		where $\beta$ is an arbitrary positive constant. Therefore,
		\begin{align*}
			\dot{V}\le\be^T\Big(\frac1N\bR_N\otimes\bF-\bL\otimes\bH^{(s)}+\frac{\beta}2\bI_{Nn}\Big)\be
			+\frac N{2\beta}\bgamma_c^T\bGamma\bgamma_c.
		\end{align*}
		Since $\bR_N$ and \bL~ are both symmetric and Laplacian, they commute, \textit{i.e.}, $\bL\bR_N=\bR_N\bL$. From Lemma \ref{Lemma: Joint Diagonalization} there exists a unitary matrix, \bQ, such that $\bR_N$ and \bL~are jointly diagonalizable 
		\begin{align*}\begin{array}{lr}
			\bL=\bQ\bJ_L\bQ^T& \bR_N=\bQ\bJ_R\bQ^T,  \end{array}
		\end{align*}
		where $\bJ_R=\mbox{diag}([0,~N,\cdots,~N])$ and $\bJ_L$ has the eigenvalues of the Laplacian $\bL$ as its diagonal entries.
		Define
					\[\bbeta\triangleq (\bQ\otimes\bI_n)\be,\]
				then
					\begin{align*}
						\dot{V}\le&\bbeta^T\left(\frac1N\bJ_R\otimes\bF-\bJ_L\otimes{\bH^{(s)}}+\frac{\beta}{2}\bI_{Nn}\right)\bbeta\\
						          &+\frac N{2\beta}\bgamma_c^T\bGamma\bgamma_c.
					\end{align*}
						
				{As any Laplacian matrix of a connected network has one  zero eigenvalue, $\bJ_R^{(N)}=\bJ_L^{(N)}=0$, with eigenvector $\bq_N=\bone_N/\sqrt{N}$; Thus,}
					\[(\bq_N^T\otimes\bI_n)\be=\sum_{i=1}^N\be_i/\sqrt{N}={\bbeta_N}.\]
				Since $\sum_{i=1}^N\be_i=\b0$, the last component of $\bbeta$ is zero, that is, $\bbeta_N=\b0$. Now, if there exists a constant $\rho>0$ such that
						\begin{align*}
							\bF-{\mu_i}\bH^{(s)}+(\rho-{\beta}/{2})\bI_{n}\prec\b0,\quad \forall i\in\{1,\,\cdots,\,N-1\}
						\end{align*}
						where $\mu_i$'s are nonzero eigenvalue of the Laplacian matrix, \bL~(if the network is connected, \bL~has $N-1$ positive eigenvalues which is a direct result of employing Assumption \ref{Assumption: Connected}); Hence,
						\begin{align*}
							\dot{V}(\be)\le -(\rho-\beta/{2})\|\be\|^2 + N\bgamma_c^T\bGamma\bgamma_c / (2\beta).
						\end{align*}
				Using lemma \ref{Lemma: Bounded x} and setting $\varepsilon^{2}= N\bgamma_c^T\bGamma\bgamma_c/(2 \beta \rho-2 \beta \epsilon-\beta^2)$, we have
						\begin{align*}
						\begin{array}{lc}
							\dot{V}(\be)\le -\epsilon\|\be\|^2, &\forall \|\be\|\ge\sqrt{N\bgamma_c^T\bGamma\bgamma_c/(2 \beta \rho-2 \beta \epsilon-\beta^2)}.\end{array}
						\end{align*}
			Define $\lambda^\star$ as \eqref{eq: lambda_star}; If $\epsilon>0$ is arbitrarily small number, $\beta=\lambda^\star-\epsilon$ maximizes the denominator subject to the stability condition, $\lambda^\star-{\beta}/{2}-\epsilon \ge 0$. Therefore,
			\begin{align*}
				\begin{array}{lc}
					\dot{V}(\be)\le -{\epsilon}\|\be\|^2, &\forall \|\be\|\ge\sqrt{ {N}{\bgamma_c^T\bGamma\bgamma_c}/{(\lambda^\star-\epsilon)}^2}.
				\end{array}
			\end{align*}
			Applying Lemma \ref{Lemma: Bounded x}, the error of the network will be bounded by \eqref{eq: Theorem1} in finite time.
	\end{IEEEproof} 
%%%%%%%%%%%%%%%%%%%%
\section{Proof of Theorem \ref{Theorem: MismatchEstimation}}\label{Proof: MismatchEstimation}
\begin{IEEEproof}
		Let 
		\[\tgamma_i\triangleq\bgamma_i-\hgamma,\] 
		\[V=\frac12\sum_{i=1}^N\be_i^T\be_i+\sum_{i=1}^N\frac1{2k_i}\tgamma_i^T\tgamma_i.\]
		Then,
		\begin{align*}
			\dot{V}=&\frac12\sum_{i=1}^N\dbe_i^T\be_i+\be_i^T\dbe_i+\sum_{i=1}^N\frac1{k_i}\dot{\tgamma}_i^T\tgamma_i\\
				   =&\sum_{i=1}^N\be_i^T[\bff(\bx_i)-\bff(\bs)]+\be_i^T\bG(\bx_i)\tgamma_i- c_i\be_i^T\bH^{(s)}\be_i\\
				    &-\sum_{i,j=1}^Nl_{ij}\be_i^T\bH^{(s)}\be_j-\sum_{i=1}^N\frac1{k_i}\dot{\hgamma}^T_i\tgamma_i,
		\end{align*}
		substituting $\dot{\hgamma}$ from \eqref{eq: input} 
		\begin{align*}
					\dot{V}=&\sum_{i=1}^N\be_i^T[\bff(\bx_i)-\bff(\bs)]+(\be_i^T\bG(\bx_i)-\frac1{k_i}\dot{\hgamma}^T_i)\tgamma_i\\
						    &-\sum_{i,j=1}^Nl_{ij}\be_i^T\bH^{(s)}\be_j- \sum_{i=1}^Nc_i\be_i^T\bH^{(s)}\be_i\\
						 \le&\be^T(\bI_N\otimes\bF-(\bL+\bC)\otimes\bH^{(s)})\be.
				\end{align*}
		If \eqref{eq: Theorem2} holds, then using Lemma \ref{Lemma: asymptotic W}, we conclude that $\|\be\| $ uniformly goes to zero, $\|\be\|\to 0$, and $\|\tgamma_i\|$'s are bounded.
		\end{IEEEproof}
		
%%%%%%%%%%%%%%%%%%%
\section{Proof of Theorem 3}\label{proof: 3}
\begin{IEEEproof}
	Substituting \eqref{eq: Distributed_input} in \eqref{eq: Error_Ref}, we have
	\begin{align}
		\dbe_i = \bff(\bx_i) - \bff(\bs) - \sum_{j = 1}^N (l_{ij}+b_{ij}) \bH \be_j - g_i \bH \be_i + \bG(\bx_i) \tilde{\bgamma}_i,\label{eq: proof0}
	\end{align}
	where $\tilde{\bgamma}_i \triangleq \bgamma_i - \hat{\bgamma}_i$. Now let $\bC = [c_{ij}]$ be the Laplacian of connected undirected graph on $1,\,\cdots,\, N$, then $\bC + \bZ^\prime$ is symmetric positive definite matrix if there exists at least one $g_i^\prime>0$ where $\bZ^\prime \triangleq \diag([z^\prime_1 \, \cdots \, z^\prime_N]^T)$ \cite{Chen07}. Hence 
	\[ V = \frac12 \sum_{i ,j =1}^N c_{ij} \be_i^T \be_j + \frac12 \sum_{i =1}^N z^\prime_i \be_i^T\be_i + \frac12\sum_{i=1}^N \tilde{\bgamma}_i^T\tilde{\bgamma}_i,\]
	is a candidate Lyapunov function.
	\begin{align}
		\dot{V} = \underbrace{\sum_{i ,j =1}^N c_{ij} \be_i^T \dbe_j  }_{\triangleq V_1}+ \underbrace{\sum_{i =1}^N z^\prime_i \be_i^T\dbe_i }_{\triangleq V_2} - \sum_{i=1}^N \tilde{\bgamma}_i^T\dot{\hat{\bgamma}}_i. \label{eq: proof1}
	\end{align}
	By substituting \eqref{eq: proof0} in the first term in \eqref{eq: proof1}, we have
	\begin{align}
		V_1 =& \sum_{i,j =1}^N c_{ij} \be_i^T (\bff(\bx_j) - \bff(\bs))- \!\! \!\! \sum_{i,j,k =1}^N c_{ij}(l_{jk}+b_{jk})\be_i^T\bH\be_k\nonumber\\
		&+\sum_{i,j =1}^N g_jc_{ij} \be_i^T\bH \be_j + \sum_{i,j =1}^Nc_{ij}\be_i\bG(\bx_j)\tilde{\bgamma}_j\nonumber\\
			{=}&  \sum_{i,j =1}^N c_{ij} \be_i^T \bff(\bx_j)-\be^T\left(\bC(\bL+\bB+\bG)\otimes\bHs \right)\be\nonumber\\
		&+ \sum_{i,j =1}^Nc_{ij}\be_i^T\bG(\bx_j)\tilde{\bgamma}_j,
	\end{align}
	where last equality is due to zero row-sum property of $\bC$. Also $\sum_{i,j =1}^N c_{ij} \be_i^T \bff(\bx_i) = \sum_{i=1}^N  \be_i^T \bff(\bx_i)\sum_{j = 1}^N c_{ij} = \b0$, hence, we have
	\begin{align*}
	\sum_{i,j =1}^N c_{ij} \be_i^T \bff(\bx_j) & = \sum_{i, \, j =1}^N c_{ij} \be_i^T \bff(\bx_j) - \sum_{i,j =1}^N c_{ij} \be_i^T \bff(\bx_i)\\
								& = \sum_{i, \, j =1}^N c_{ij} \be_i^T\Big( \bff(\bx_j) - \bff(\bx_i) \Big)\\
								& = \frac12 \sum_{i,\, j =1}^N c_{ij} \be_i^T \Big( \bff(\bx_j) - \bff(\bx_i) \Big) \\
								&\quad \quad +\frac12 \sum_{i,j =1}^N c_{ij} \be_i^T \Big( \bff(\bx_j) - \bff(\bx_i) \Big) \\
								& = \frac12 \sum_{i,\, j =1}^N c_{ij} \be_i^T \Big( \bff(\bx_j) - \bff(\bx_i) \Big) \\
								&\quad \quad +\frac12 \sum_{i,j =1}^N c_{ji} \be_j^T \Big( \bff(\bx_i) - \bff(\bx_j) \Big)
\\
								&= \frac12 \sum_{i,\, j =1}^N c_{ij} (\be_i - \be_j)^T \Big( \bff(\bx_i) - \bff(\bx_j) \Big) \\
								& \le \frac12 \sum_{i,\, j =1}^N |c_{ij}|(\be_i - \be_j)^T \bF (\be_i- \be_j) \\
		& \le \sum_{\tiny\begin{array}{l} i,\,j =1, \\ j\ne \,  i\end{array}}^N \!\! \!\!  |c_{ij}| ~\be_i ^T\,  \bF \, \be_i - \!\! \!\! \!\! \sum_{\tiny\begin{array}{l} i,\,j =1, \\ j\ne \,  i\end{array}}\!\! \!\!  |c_{ij}| \, \be_i^T \, \bF \, \be_j 
	\end{align*}
	\begin{align*}
								& \le \sum_{i=1}^N ~\be_i \,  \bF \, \be_i \left(\sum_{\tiny\begin{array}{l} j =1, \\ j\ne \,  i\end{array}}^N \!\! \!\! |c_{ij}| \right)\\
								& \quad \quad- \, \sum_{\tiny\begin{array}{l} i,\,j =1, \\ j\ne \,  i\end{array}}|c_{ij}| \, \be_i  \, \bF \, \be_j \\
								& \le \be^T \left(\bC \otimes \bF \right) \be.
	\end{align*}
	Hence,
	\begin{align}
	V_1 \le ~ & \be^T\left(\bC \otimes \bF - \bC(\bL+\bB+\bG)\otimes\bHs\right)\be\nonumber\\
		&+ \sum_{i,j =1}^Nc_{ij}\be_i^T\bG(\bx_j)\tilde{\bgamma}_j. 
	\end{align}
	Therefore,
	\begin{align}
	\dot{V} \le \be^T\left((\bC+\bZ^\prime) \otimes \bF - (\bC+\bZ^\prime)(\bL+\bB+\bG)\otimes\bHs \right)\be \nonumber\\
	+\sum_{i =1}^Nz^\prime_{i}\be_i^T\bG(\bx_j)\tilde{\bgamma}_j + \sum_{i,j =1}^Nc_{ij}\be_i^T\bG(\bx_j)\tilde{\bgamma}_j - \sum_{i =1}^N \dot{\hat{\bgamma}}_i^T\tilde{\bgamma}_i.\nonumber
	\end{align}
	Substituting $\dot{\hat{\bgamma}}_i$ from \eqref{eq: Distributed_Estimation}, yields
	\begin{align*}
	\dot{V} \le \be^T \!\!  \left([(\bC+\bZ^\prime) \otimes \bI_n ]\!\! \left[\bI_N \otimes \bF - (\bL+\bB+\bZ)\otimes\bHs \right]\right)\be.
	\end{align*}
	 If conditions in \eqref{eq: Theorem3} hold, Lemma \ref{Lemma: asymptotic W} implies that $\be_i$'s are asymptotically stable and $\tilde{\bgamma}_i$'s are bounded. It should be noted that the product of two symmetric positive definite matrices is a positive definite matrix (not necessarily symmetric).This can be shown by the Weyl's inequalities for product of two symmetric matrices.
	\end{IEEEproof}

\bibliographystyle{IEEETran}
\bibliography{mybib}

\end{document}